# Sensitivity mapping of TBL wall-pressure spectra with CFD turbulence models for wind tunnel test result prediction

Biplab Ranjan Adhikary[1], Ananya Majumdar[1], Subhadeep Sarkar[2], and Partha Bhattacharya[1]

[1]Department of Civil Engineering, Jadavpur University, Kolkata-700032, India
[2]Department of Civil Engineering, IIEST Shibpur, Howrah-711103, India

**ABSTRACT**

In the present work, an attempt is made to map the sensitivity of the existing zero pressure gradient (ZPG) turbulent boundary layer (TBL) wall-pressure spectrum models with different TBL parameters, and eventually, with different Reynolds Averaged Navier Stokes (RANS) turbulence models, simulated in OpenFOAM and ANSYS Fluent solvers. This study will help future researchers to choose a particular RANS turbulence model vis-à-vis a particular wall-spectrum model in order to obtain a reasonably accurate wind tunnel result predicting capability. First, the best-predicting pressure spectrum models are selected by comparing them with wind tunnel test data. Next, considering the experimental TBL parameters as benchmarks, errors in RANS-produced data are estimated. Furthermore, wall-pressure spectra are calculated following semi-empirical spectrum models using TBL parameter feed obtained from experiments and computational fluid dynamics (CFD) simulations. Finally, sensitivity mapping is performed between spectrum models and the RANS models, with different normalized wall-normal distances ($y^+$).

## 1. INTRODUCTION

Turbulent boundary layer (TBL) wall-pressure fluctuation is one of the key parameters for vibro-acoustic response prediction of TBL-excited flexible panels, mostly in aircraft, automobile, and marine industries. There are several semi-empirical single-point wall-pressure spectrum models [1-6] available and widely used for practical purposes. However, these models are essentially dependent on experimental data feeding. Now, as the experiments are quite expensive, time and resource-consuming, there are intense searches for alternate methods to the experimentations. With the increasing computing capacity, CFD simulations have emerged to fill the gap. Although direct numerical simulation (DNS) and large eddy simulation (LES) are proven to be robust and more accurate, considering the computation cost RANS techniques [7] are most commonly used for industrial purposes nowadays.

However, all the RANS models do not provide a similar level of accuracy for predicting TBL parameters. Their accuracy changes from case to case. Change in $y^+$ values or even change in the solver alters the accuracy. Moreover, only the universal velocity plot ($y^+$ vs $U^+$) cannot be the determining factor for the prediction of the wall-pressure fluctuations. On the other hand, due to the differences in the method of their development, wall-spectrum models are differently sensitive to the TBL parameters. Therefore, extensive research work is required to a) identify the best spectrum model to predict wind tunnel results, b) quantify the accuracy of different solvers, CFD models and $y^+$ values in comparison to the experimental results, c) proper sensitivity mapping of the spectrum models with used CFD utilities. The present study will definitely serve as comprehensive documentation for future researchers in the said domain.

## 2. LITERATURE REVIEW AND OBJECTIVE

In one of the latest works, Thomson and Rocha [8] compared different wind tunnel and in-flight test results with available semi-empirical spectrum models for zero pressure gradient cases. They have found Goody [1] and Smol'yakov [2] models to be the best predictor to the wind tunnel test results. Leneveu et al. [9] performed both wind tunnel test and CFD simulation using OpenFOAM solver. Dominique et al. [10] used an artificial neural network (ANN) that is trained with existing experimental and CFD results and predicts better spectrum formulation, especially for the adverse pressure gradients.

However, an attempt to map the best predicting spectrum models for wind tunnel test data with the in-house simulated CFD models with possible variants is non-existing. Therefore, in the present work, a careful study is carried out involving a) identification of the best predicting single-point wall-pressure spectrum model with regard to wind tunnel experiment, b) quantification of accuracy of OpenFOAM and Fluent (V14.5) solvers, different CFD-RANS models with varying $y^+$ values, c) mapping the sensitivity of the spectrum models with varied CFD simulation and utility.

## 3. METHODOLOGY

In the present work, at first wall-pressure spectra are calculated using several semi-empirical models with experimental TBL parameters. Out of them, Goody and Smol'yakov models are found to be the best predictor in accordance with the wind tunnel test results (Fig. 1), which is in good agreement with the findings of Thomson and Rocha [8].

Mathematical descriptions of TBL pressure spectra proposed by Goody [1] and Smol'yakov [2] are provided in the following equations:

Goody model

$$\Phi_P = \frac{3(\omega\tau_\omega)^2(\frac{\delta}{U})^3}{(\left(\frac{\omega\delta}{U}\right)^{0.75} + 0.5)^{3.7} + (1.1R_T^{-0.57}\frac{\omega\delta}{U})^7} \quad (1)$$

Where: $R_T = 0.11(\frac{U_e\theta}{\nu})^{3/4}$ (2)

Smol'yakov model:

$$\Phi_P = \frac{3(\omega\tau_\omega)^2(\frac{\delta}{U})^3}{(\left(\frac{\omega\delta}{U}\right)^{0.75} + 0.5)^{3.7} + (1.1R_T^{-0.57}\frac{\omega\delta}{U})^7} \quad (3)$$

$$\Phi_P = \frac{1.49}{u_\tau^2}10^{-5}\tau_\omega^2 \nu Re_\theta^{2.74}\bar{f}^2\left(1 - 0.117Re_\theta^{0.44}\bar{f}^{\frac{1}{2}}\right) \quad (4)$$
$$\text{for } \bar{f} > \bar{f}_0$$

$$\Phi_P = \frac{2.75\,\tau_\omega^2\nu}{u_\tau^2\,\bar{f}^{-1.11}}\left(1 - 0.82e^{-0.51\left(\frac{\bar{f}}{\bar{f}_0}-1\right)}\right) \quad (5)$$
$$\text{for } \bar{f}_0 < \bar{f} < 0.2$$

$$\Phi_P = \frac{\tau_\omega^2\nu[38.9e^{-8.35\bar{f}} + 18.6e^{-3.58\bar{f}} + 0.31e^{-2.14\bar{f}}]}{1 - 0.82e^{-0.51\left(\frac{\bar{f}}{\bar{f}_0}-1\right)}} \quad (6)$$
$$\text{for } \bar{f} > 0.2$$

Where: $\bar{f} = \frac{2\pi f\nu}{U_\tau^2}$; $\bar{f}_0 = 49.35Re_\theta^{-0.88}$ (7)

Other spectrum models used in this work are detailed in [3-6].

Next, a flat plate TBL wind tunnel experiment conducted by Leneveu et al. [9] is simulated using *four* RANS turbulence models [7]:
i) Standard $k-\omega$  ii) $k-\omega$ SST
iii) $k-\epsilon$  iv) Realizable $k-\epsilon$

Mesh sensitivity studies for the flat plate boundary layer case [9] are first performed separately for OpenFOAM and Fluent solvers for different RANS models. Next, these two solvers are employed to simulate the flow field and extract the TBL parameters in alignment with the experiment at two different locations (point 1: 1495mm and point 2: 1595mm downstream of the domain inlet) and two different wind speeds (30m/s and 50m/s). Then, different CFD models with different $y^+$ values are compared with experimental results both at the component level $(U_\tau, \delta, \delta^*, \theta)$ and pressure spectrum level $(\Phi_P)$. Finally, sensitivity mapping is performed between a particular spectrum model and a particular CFD setup and presented.

## 4. RESULTS AND DISCUSSION

This section contains *three* subsections:
4.1) Comparison of single-point wall-pressure spectrum models with wind tunnel experimental PSD result.
4.2) Mesh sensitivity study of CFD models.
4.3) Accuracy analysis of different solvers, RANS models, and $y^+$ values. Sensitivity mapping.

### 4.1 Comparison of spectrum models with wind tunnel PSD

Several semi-empirical spectrum models are compared with PSD values from wind tunnel experiments conducted by Goody and Simpson [10], and presented in Fig. 1. The experimental spectrum is single-sided, for a two-dimensional ZPG flow [8]. They used naturally developed turbulent boundary layers to study wall-pressure fluctuations. Among all the spectrum models Goody and Smol'yakov models are found to be the best in predicting the experimental results, as per Fig. 1.

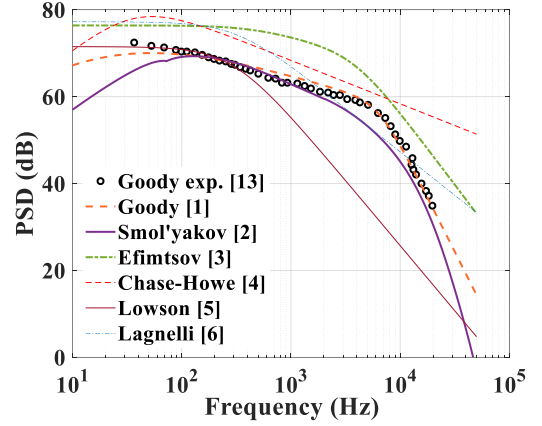

**Figure 1:** Comparison of wall-spectrum models with Goody-Simpson experiment [13]. Ref. $4 \times 10^{-10}$ Pa$^2$

### 4.2 CFD: Mesh sensitivity study

Three types of meshing with uniform rectangular cells are studied for both OpenFOAM and Fluent solvers and presented in Table 1. The representative cell length (h) is calculated as, $h = \frac{1}{N}\sum_{cell} A_p^{1/2}$, where N is the number of cells and $A_p$, the area of each cell.

**Table 1: Three different 2D meshes**

| Mesh | Number of cells | Representative Cell Length (h) [m] |
|---|---|---|
| Coarse | 60000 | 0.0029 |
| Medium | 120000 | 0.0021 |
| Fine | 240000 | 0.0015 |

The present meshing satisfies the recommendations of Celik et al. [11] that the representative cell lengths should be at least 30% different for each mesh. The mesh sensitivity study is performed by referencing the friction velocity value obtained from the experiment conducted by Leneveu et al. [9] and presented in Fig. 2 to Fig. 5.

The most ideal mesh is having an infinite number of cells, corresponding to a representative cell length (RCL) (h) of *zero*. On designating fine, medium, and coarse mesh as 1, 2, and 3, respectively, friction velocity at $h = 0$ can be expressed as,

$$U_{\tau 0} = \frac{r_{21}^p U_{\tau 1} - U_{\tau 2}}{r_{21}^p - 1} \quad (8)$$

where refinement ratio, $r_{21} = h_2/h_1$, and p is the order of convergence. The order of convergence p is estimated using the method proposed by Celik et al. [11], since it is a more general approach suitable for both monotonic and oscillatory convergence. In the beginning differences in calculated friction velocity between the fine mesh and the medium mesh ($\epsilon_{21}$), and between the medium mesh and the coarse mesh ($\epsilon_{32}$) are determined as:

$$\epsilon_{21} = U_{\tau 2} - U_{\tau 1}; \epsilon_{32} = U_{\tau 3} - U_{\tau 2} \quad (9)$$

Next, their ratio is calculated to find $s = sign(\frac{\epsilon_{32}}{\epsilon_{21}})$

Final form of the implicit non-linear equation is:

$$\frac{1}{\ln(r_{21})}\left|\ln|\epsilon_{32}/\epsilon_{31}| + \ln\left(\frac{r_{21}^p - s}{r_{32}^p - s}\right)\right| - p = 0 \quad (10)$$

This equation is solved using the Newton-Raphson iteration technique in MS Excel.

In the present mesh sensitivity study, *three* types of errors are estimated for each case; relative error ($e_{21}$), extrapolated relative error ($e_{21}^{ext}$), and Grid Convergence Index ($GCI_{21}$).

$$e_{21} = \left|\frac{U_{\tau 2} - U_{\tau 1}}{U_{\tau 1}}\right|; e_{21}^{ext} = \left|\frac{U_{\tau 1} - U_{\tau 0}}{U_{\tau 0}}\right|; GCI_{21} = \left|\frac{e_{21}}{r_{21}^p - 1}\right| \quad (11)$$

Only *four* of them are presented in Table 2 to Table 5 as representative results. Detailed studies are presented in Fig. 2 to Fig. 5.

**Table 2: Point 1; $k - \omega$ SST; OpenFOAM; $U_\infty = 30 m/s$**

| RCL (h) [m] | $U_\tau$ [m/s] | $U_\tau$ extr [m/s] | $e_{21}$ [%] | $e_{21}^{ext}$ [%] | $GCI_{21}$ [%] |
|---|---|---|---|---|---|
| 0.0015 | 1.156 | | | | |
| 0.0021 | 1.154 | 1.159 | 0.19 | 0.29 | 0.36 |
| 0.0029 | 1.159 | | | | |

**Table 3: Point 1; $k - \omega$ SST; Fluent; $U_\infty = 30 m/s$**

| RCL (h) [m] | $U_\tau$ [m/s] | $U_\tau$ extr [m/s] | $e_{21}$ [%] | $e_{21}^{ext}$ [%] | $GCI_{21}$ [%] |
|---|---|---|---|---|---|
| 0.0015 | 1.140 | | | | |
| 0.0021 | 1.140 | 1.140 | 0.00 | 0.00 | 0.00 |
| 0.0029 | 1.150 | | | | |

**Table 4: Point 2; $k - \omega$ SST; OpenFOAM; $U_\infty = 30 m/s$**

| RCL (h) [m] | $U_\tau$ [m/s] | $U_\tau$ extr [m/s] | $e_{21}$ [%] | $e_{21}^{ext}$ [%] | $GCI_{21}$ [%] |
|---|---|---|---|---|---|
| 0.0015 | 1.151 | | | | |
| 0.0021 | 1.148 | 1.156 | 0.22 | 0.48 | 0.60 |
| 0.0029 | 1.153 | | | | |

**Table 5: Point 2; $k - \omega$ SST; Fluent; $U_\infty = 30 m/s$**

| RCL (h) [m] | $U_\tau$ [m/s] | $U_\tau$ extr [m/s] | $e_{21}$ [%] | $e_{21}^{ext}$ [%] | $GCI_{21}$ [%] |
|---|---|---|---|---|---|
| 0.0015 | 1.130 | | | | |
| 0.0021 | 1.140 | 1.120 | 0.88 | 0.93 | 1.15 |
| 0.0029 | 1.150 | | | | |

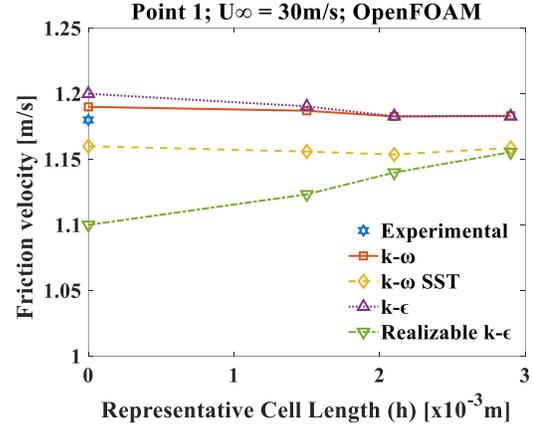

**Figure 2: Mesh sensitivity study; OpenFOAM**

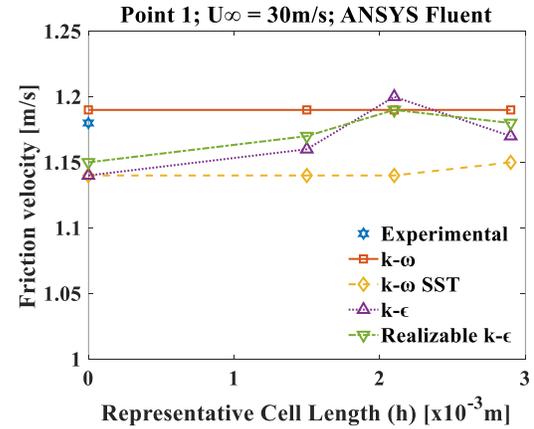

**Figure 3: Mesh sensitivity study; Fluent**

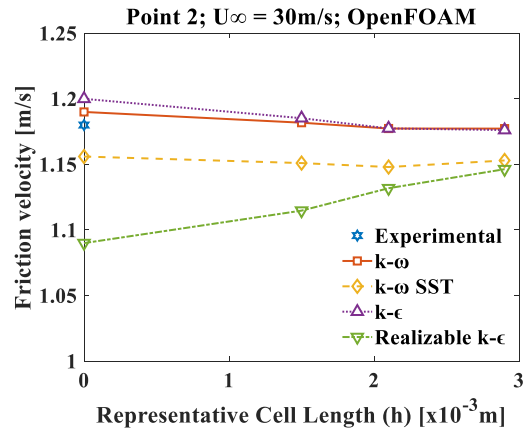

**Figure 4: Mesh sensitivity study; OpenFOAM**

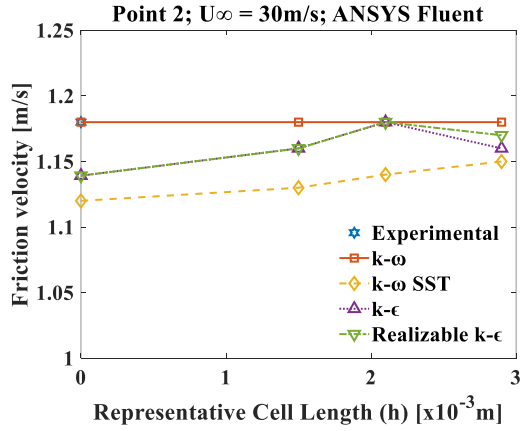

**Figure 5:** Mesh sensitivity study; Fluent

As observed from the mesh sensitivity studies, it is found that the 'medium' mesh is sufficient for the subsequent studies, and thus the accuracy analysis of different CFD variants is conducted with this meshing. Different solvers (OpenFOAM, Fluent), turbulence models ($k-\omega$, $k-\omega$ SST, $k-\epsilon$ and Realizable $k-\epsilon$), $y^+$ values (1, 30 and 100) are examined with the experimental values ($U_\tau, \delta, \delta^*, \theta$) obtained [12].

### 4.3 CFD simulation; Part 1: Clustering and convergence

Mathematical formulation of the RANS models can be found in reference [7]. Here, the formulation for the normalized wall distance and the numerical technique used to estimate $\delta^*$ and $\theta$ from the CFD-obtained velocity profiles.

Normalized wall distance:

$$y^+ = \frac{y_p U_\tau}{\nu} \quad (12)$$

Where: $y_p$ is the distance of the centroid of the first cell adjacent to the wall.

Displacement thickness ($\delta^*$) and momentum thickness :

$$\delta^* = \sum_{i=2}^{N-1} (1 - \frac{u_i}{U_0})(\frac{y_{i+1} - y_{i-1}}{2}) \quad (13)$$

$$\theta = \sum_{i=2}^{N-1} \frac{u_i}{U_0}(1 - \frac{u_i}{U_0})(\frac{y_{i+1} - y_{i-1}}{2}) \quad (14)$$

In order to simulate wall parameters properly, clustering near-wall is performed. For different $y^+$ values first cell height ($2y_p$) is initially calculated as per Eq. 12 and are presented in Table 6 and Table 7. Bias Factor (BF) is the ratio of the last cell height to the first cell height. Fig. 7 and Fig. 8 depict universal velocity plots.

**Table 6:** $y^+$ calculation for $U_\infty = 30 m/s$

| $y^+$ | $U_\tau$ [m/s] | $\nu$ [$10^{-5}$ m²/s] | $2y_p$ [$10^{-5}$ m] | BF |
|---|---|---|---|---|
| 1 | | | 2.47 | 284.3 |
| 30 | 1.18 | 1.46 | 74.5 | 2.6 |
| 100 | | | 247 | 1 |

**Table 7:** $y^+$ calculation for $U_\infty = 50 m/s$

| $y^+$ | $U_\tau$ [m/s] | $\nu$ [$10^{-5}$ m²/s] | $2y_p$ [$10^{-5}$ m] | BF |
|---|---|---|---|---|
| 1 | | | 1.55 | 496.29 |
| 30 | 1.89 | 1.46 | 46.5 | 5.64 |
| 100 | | | 155 | 1 |

Convergence shown for the realizable $k-\epsilon$ model with $y^+$ 30.

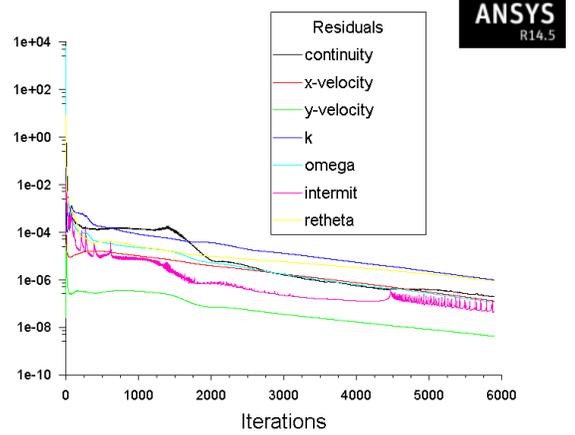

**Figure 6:** Convergence; $U_\infty = 30 m/s$; $y^+ = 30$; Fluent

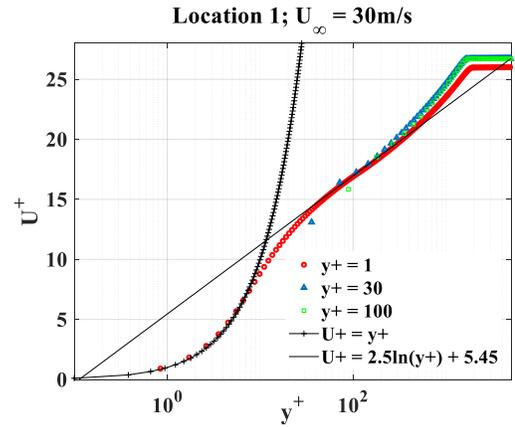

**Figure 7:** Universal velocity plot; Fluent

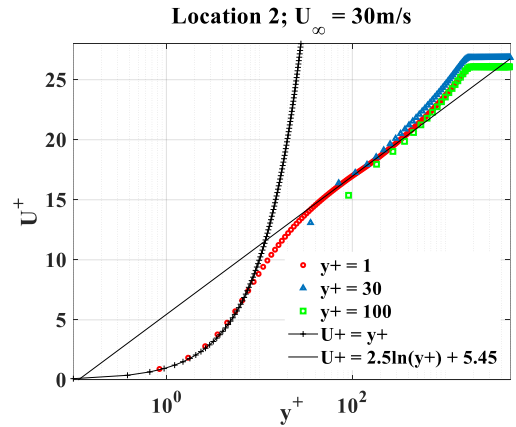

**Figure 8:** Universal velocity plot; Fluent

## 4.4 CFD simulation; Part 2: Component error analysis

In this segment, accuracy of different turbulence models with CFD models for various $y^+$ values are estimated in terms of $U_\tau, \delta, \delta^*$ and $\theta$. The experimental TBL parameters are set as reference.

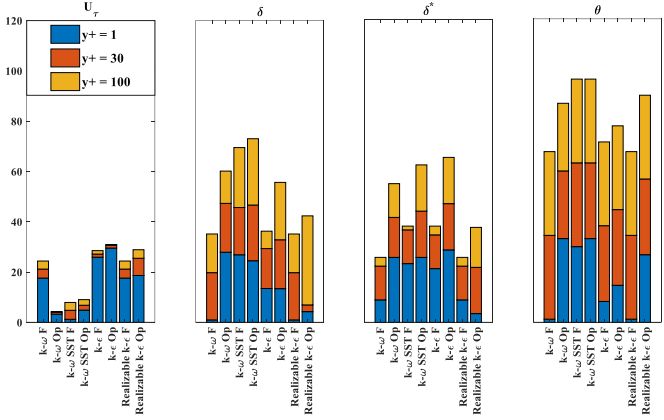

Figure 9: Component error (%); $U_\infty = 30$ m/s; Point 1

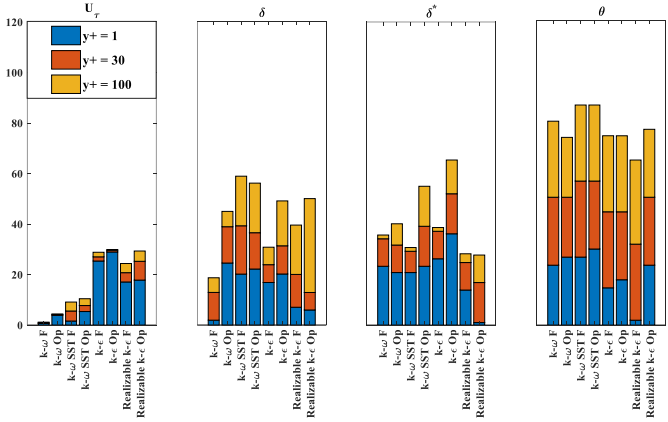

Figure 10: Component error (%); $U_\infty = 30$ m/s; Point 2

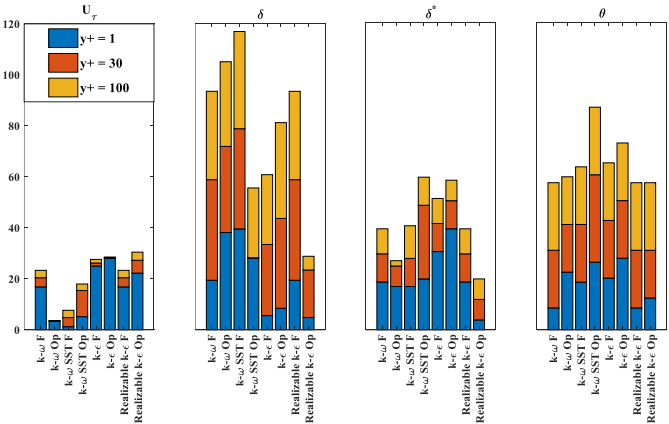

Figure 11: Component error (%); $U_\infty = 50$ m/s; Point 1

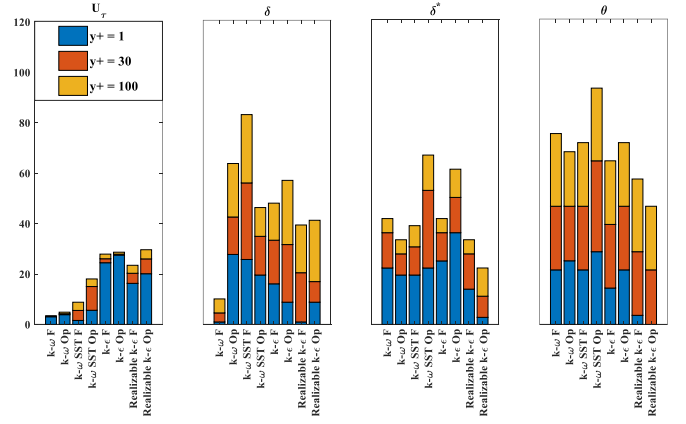

Figure 12: Component error (%); $U_\infty = 50$ m/s; Point 2

## 4.5 CFD simulation; Part 3: $\bar{p}^2$ error analysis

Here, at first the single-sided wall pressure spectra ($\Phi_p$) are calculated using Goody and Smol'yakov models for the CFD-obtained TBL parameters. Next, one can sum the spectrum over its collapsing frequency (50 kHz for the present case) as per Eq. 15 and estimate the mean square of pressure fluctuations ($\bar{p}^2$) and compare them with the experimental $\bar{p}^2$ values.

$$\bar{p}^2 = \int_0^\infty \Phi_p(f)\, df \quad (15)$$

This is a practical approach as it accounts for the global energy and not the local pressure PSD values.

Table 7: Error analysis of $\bar{p}^2$; Goody model; Point 1

| Model | $y^+$ | 30m/s | | 50m/s | |
|---|---|---|---|---|---|
| | | Open FOAM | Fluent | Open FOAM | Fluent |
| $k-\omega$ | 1 | 0.5 | 0.4 | -1.1 | 1.0 |
| | 30 | 1.4 | 0.3 | **-0.1** | **0.3** |
| | 100 | 0.9 | 0.4 | **-0.1** | 1.0 |
| $k-\omega$ SST | 1 | **-0.2** | 1.1 | -1.4 | -0.5 |
| | 30 | 0.7 | **0.0** | 2.6 | -1.2 |
| | 100 | 0.8 | 0.4 | -0.7 | -0.5 |
| $k-\epsilon$ | 1 | 7.1 | 6.2 | 6.8 | 6.1 |
| | 30 | 1.5 | 0.6 | **-0.1** | **-0.3** |
| | 100 | 1.4 | **0.0** | -0.3 | 6.1 |
| Realizable $k-\epsilon$ | 1 | 5.1 | 5.0 | 5.5 | 4.3 |
| | 30 | -2.1 | **0.0** | -1.3 | -1.2 |
| | 100 | -4.2 | **0.0** | -1.0 | 4.3 |

TBL pressure PSD using Goody model can be estimated using Fluent ($k-\omega$ SST, $k-\epsilon$ and Realizable $k-\epsilon$) solver even with 0% error. OpenFOAM predicts the same with -0.2% error. With the increase in flow velocity, the accuracy of the OpenFOAM solver increases.

**Table 8: Error analysis of $\bar{p}^2$; Smol'yakov model; Point 1**

| Model | $y^+$ | 30m/s | | 50m/s | |
|---|---|---|---|---|---|
| | | Open FOAM | Fluent | Open FOAM | Fluent |
| $k-\omega$ | 1 | **-0.4** | 1.6 | -2.1 | 1.9 |
| | 30 | 1.5 | **1.4** | **0.0** | **0.3** |
| | 100 | **0.4** | **1.4** | -0.1 | **0.3** |
| $k-\omega$ SST | 1 | -3.7 | -1.7 | -3.4 | -0.7 |
| | 30 | -3.2 | -6.4 | 5.7 | -2.3 |
| | 100 | -4.1 | -6.2 | -1.7 | -1.8 |
| $k-\epsilon$ | 1 | 69.1 | 36.8 | 13.4 | 12.2 |
| | 30 | 1.6 | -1.7 | **0.0** | -0.7 |
| | 100 | **0.4** | -2.9 | -0.2 | -0.9 |
| Realizable $k-\epsilon$ | 1 | -3.3 | 26.9 | 11.1 | 8.8 |
| | 30 | -13.1 | -6.4 | -3.4 | -2.4 |
| | 100 | -6.3 | -6.2 | -2.0 | -1.8 |

In case of Smol'yakov model with higher flow speed $k-\omega$ and $k-\epsilon$ models in OpenFOAM solver achieve 100% accuracy. $k-\omega$ model turns out to be best option in Fluent solver for both the flow speeds.

## 5. CONCLUSIONS

The conclusions of the present research work are as follows:
a) Wind tunnel experimental results for ZPG flat plate TBL cases are best predicted by Goody and Smol'yakov single-point pressure spectrum models.
b) All the comments on CFD turbulence models are restricted to flat plate cases; wind tunnel experiments.
c) Goody and Smol'yakov models are mapped with the specific CFD-RANS model utilities that one can adopt in future. The mapping is done through TBL parameters, which alters the sensitivity of any model. As the TBL parameters are coupled in the semi-empirical equations, it is only possible to predict the final error in $\bar{p}^2$ with extensive CFD analysis.
d) Even with significant errors in component (TBL parameters) level one should not stop the study as the final prediction of the wall-pressure fluctuations can be shockingly accurate, producing even *ZERO* errors.

| Model/ solver | $y^+$ | Error (%) | | | |
|---|---|---|---|---|---|
| | | $U_\tau$ | $\delta$ | $\delta^*$ | $\bar{p}^2$ |
| $k-\omega$ SST (Fluent) | 30 | -3.57 | -18.79 | -13.37 | 0.0 |
| $k-\epsilon$ (OpenFOAM) | 30 | 0.07 | -35.17 | -10.98 | 0.0 |

e) As a general observation, 30 is found to be the most suited value for $y^+$ in terms of $\bar{p}^2$ prediction accuracy.
f) $k-\epsilon$ family models with $y^+=1$ *MUST* be avoided, despite the fact that the $y^+=1$ exhibits the best normalized velocity prediction in $y^+$ vs $U^+$ plots.

## NOMENCLATURE

| | | |
|---|---|---|
| $\Phi_p$ | Wall pressure PSD | [Pa$^2$/Hz] |
| $\bar{p}^2$ | Mean square pressure | [Pa$^2$] |
| $U_\tau$ | Friction velocity | [m$^2$/s] |
| $\delta$ | Boundary layer thickness | [m] |
| $\delta^*$ | Displacement thickness | [m] |
| $\theta$ | Momentum thickness | [m] |
| $\rho$ | Density of air | [kg/m$^3$] |
| $y^+$ | Normalized wall-normal distance | -- |
| $U^+$ | Normalized velocity | -- |
| $\omega$ | Radial frequency | Rad/s |
| $f$ | Cyclic frequency | Hz |